\pdfoutput=1
\RequirePackage{ifpdf}
\ifpdf % We are running pdfTeX in pdf mode
\documentclass[pdftex]{sigma}
\else
\documentclass{sigma}
\fi

\numberwithin{equation}{section}

\renewcommand\ge{\geqslant}

\renewcommand\le{\leqslant}

\newcommand\abs[1]{\left|#1\right|}

\newcommand{\tr}{\mathop{\hbox{\rm tr}}\nolimits}

\begin{document}

\renewcommand{\thefootnote}{$\star$}

\renewcommand{\PaperNumber}{078}

\FirstPageHeading

\ShortArticleName{Integrable Hierarchy of the Quantum Benjamin--Ono Equation}

\ArticleName{Integrable Hierarchy of the\\ Quantum Benjamin--Ono Equation\footnote{This paper is a~contribution to the Special Issue in honor of
Anatol Kirillov and Tetsuji Miwa.
The full collection is available at
\href{http://www.emis.de/journals/SIGMA/InfiniteAnalysis2013.html}{http://www.emis.de/journals/SIGMA/InfiniteAnalysis2013.html}}}

\Author{Maxim~NAZAROV and Evgeny~SKLYANIN}

\AuthorNameForHeading{M.~Nazarov and E.~Sklyanin}

\Address{Department of Mathematics, University of York, York YO10 5DD, United Kingdom}

\ArticleDates{Received September 26, 2013, in f\/inal form December 03, 2013; Published online December 07, 2013}

\Abstract{A hierarchy of pairwise commuting Hamiltonians for the quantum periodic Benjamin--Ono equation is
constructed by using the Lax matrix.
The eigenvectors of these Hamiltonians are Jack symmetric functions of inf\/initely many variables
$x_1,x_2,\ldots$.
This construction provides explicit expressions
for the Hamiltonians in terms of the power sum
symmetric functions $p_n=x_1^n+x_2^n+\cdots$
and is based on our recent results from~[\textit{Comm. Math. Phys.} \textbf{324} (2013), 831--849].
}


\Classification{33D52; 05E05; 37K10; 81Q80}

\renewcommand{\thefootnote}{\arabic{footnote}}
\setcounter{footnote}{0}

\section{Introduction}

Take the ring $\mathbb C[p_1,p_2,\ldots]$ of polynomials in countably many variables $p_1, p_2,\ldots$ and
let $\alpha\in\mathbb C$ be a~parameter.
For $n=1,2,\ldots$ denote by $p_n^{\ast}$ the operator
$\alpha n\partial/\partial p_n$ of renormalized dif\/ferentiation relative to the
variable $p_n$.
The following dif\/ferential operator on $\mathbb C[p_1,p_2,\ldots]$ arises in various problems of
algebra and mathematical physics and has recently enjoyed considerable attention:
\begin{gather}
\label{def-I}
\sum_{m,n\ge1}\big(p_mp_np_{m+n}^{\ast}+p_{m+n}p_m^{\ast}p_n^{\ast}\big)+(\alpha-1)\sum_{n\ge1}np_np_n^{\ast},
\end{gather}
see~\cite{AMOS,CJ,NS1,OP,Pol} for instance.
In the classical limit $\alpha\rightarrow0$, the parameter $\alpha$ playing the role of the Planck
constant, the operator~\eqref{def-I} becomes the Hamiltonian of the periodic Benjamin--Ono equation,
a~well-studied integrable Hamiltonian system~\cite{AFSS,B,KLM,KM,O}.

The ring $\mathbb C[p_1,p_2,\ldots]$ can be identif\/ied with the ring of symmetric functions in the
variables $x_1,x_2,\ldots$ by setting $p_n=x_1^{ n}+x_2^{ n}+\cdots$ for
every $n$.
It is known that then the operator~\eqref{def-I} is diagonalised in the basis of the Jack symmetric
functions corresponding to the parameter~$\alpha$.
It is also known~\cite{SV} that~\eqref{def-I} can be included into an inf\/inite family of commuting
operators, all diagonal in the basis of Jack symmetric functions.
In~\cite{NS1} we provided an explicit expression for a~generating function $A(u)$ of those commuting
operators, see Theorem~\ref{Theorem1} in Section~\ref{Section4} of the present article.
The function $A(u)$ has been def\/ined as the projective limit of the renormalized Sekiguchi--Debiard
determinant for the $N$-particle Calogero--Sutherland model as $N\rightarrow\infty$.

The purpose of the present article is to construct a~generating function $I(u)$ of another family
$I^{(1)},I^{(2)},\ldots$ of commuting operators diagonal in the basis of
the Jack symmetric functions.
Our construction is drastically dif\/ferent from that of~\cite{NS1} and has been inspired by the Lax
formulation of the classical Benjamin--Ono equation.
Namely, put
\begin{gather}
\label{iu}
I(u)=I^{(1)}u^{-1}+I^{(2)}u^{-2}+\cdots,
\end{gather}
where by def\/inition
\begin{gather}
\label{hdef}
1-u^{-1}I(u)=A(u)/A(u+1).
\end{gather}
Our main result is Theorem~\ref{Theorem2} in Section~\ref{Section4} giving an explicit formula for $I(u)$.
In particular $I^{(1)}$ is the weight counting operator
$p_1p_1^{\ast}+p_2p_2^{\ast}+\cdots$ while $I^{(2)}$
coincides with the operator~\eqref{def-I}.

An advantage of this new generating function $I(u)$ is that it is written explicitly in terms of the
collective variables $p_1,p_2,\ldots$ whereas $A(u)$ is written in terms of the monomial symmetric
functions, which are expressed in these variables in a~rather complicated combinatorial way.

Here is the plan of this article.
In Section~\ref{Section2}, following a~brief discussion of the known results on the classical Benjamin--Ono
equation, we explain our motivation and the origin of the formula~\eqref{hua}.
In Sections~\ref{Section3} and~\ref{Section4} we recall the basic facts
about symmetric functions that we need for our proofs.
In Section~\ref{Section5} we outline the results of~\cite{NS1} and describe the generating function~$A(u)$.
In Section~\ref{Section6} we state our main result, Theorem~\ref{Theorem2}.
In Section~\ref{Section7} we introduce our main tool, the Baker--Akhiezer vector~\eqref{psiu}.
In Section~\ref{Section8} we complete the proof.
Its main idea is to write~\eqref{psiu} in terms of the monomial symmetric functions similarly to $A(u)$,
see Theorem~\ref{Theorem3} in Section~\ref{Section7}.

In this article we generally keep to the notation of the book~\cite{M} for symmetric functions.
When using results from~\cite{M} we will simply indicate their numbers within the book.
For example, the statement (6.19) from Chapter I of the book will be referred to as [I.6.9] assuming it is
from~\cite{M}.

\section[Classical periodic Benjamin-Ono equation]{Classical periodic Benjamin--Ono equation}\label{Section2}

Let $s\in\mathbb R$ be a~real variable.
The classical Benjamin--Ono equation~\cite{AFSS,B,KLM,KM,O} is formulated in terms of the real-valued
f\/ield $\theta(s)$ on the line $\mathbb R$ with the Poisson bracket
\begin{gather}
\label{phi-pb}
\{\theta(s),\theta(t)\}=2\pi\delta^{\prime}(s-t).
\end{gather}

Imposing the periodicity condition $s\equiv s+2\pi$ and setting $z=\mathrm{e}^{-\mathrm{i}
s}$ one can identify $\mathbb R/2\pi\mathbb Z$ with the unit circle $S^1=\{z\in\mathbb C:\abs{z}=1\}$.
Respectively, set $\varphi(z)=\theta(s)$.
Let $p_0$ and $p_n$, $\bar{p}_n$ be the Fourier coef\/f\/icients of the function~$\varphi(z)$:
\begin{gather*}
\varphi(z)=p_0+\sum_{n\ge1}\big(p_nz^{-n}+\bar{p}_nz^n\big).
\end{gather*}
In terms of these coef\/f\/icients the Poisson bracket~\eqref{phi-pb} reads as follows:
\begin{gather*}
\{p_m,p_n\}=\{\bar{p}_m,\bar{p}_n\}=0
\qquad
\text{and}
\qquad
\{\bar{p}_m,p_n\}=\mathrm{i}n\delta_{mn}
\qquad
\text{for}
\quad
m,n\ge1;
\\
\{p_0,p_n\}=\{p_0,\bar{p_n}\}=0
\qquad
\text{for}
\quad
n\ge0.
\end{gather*}
Since $p_0$ belongs to the center of the Poisson bracket, and since the Hamiltonians we are interested in
do not depend on $p_0$, one can safely set $p_0=0$.
We will do so henceforth.

Introduce the projections $\varphi\mapsto\varphi_\pm$ where
\begin{gather*}
\varphi_+(z)=\sum_{n\ge1}\bar{p}_nz^n
\qquad
\text{and}
\qquad
\varphi_-(z)=\sum_{n\ge1}p_nz^{-n}.
\end{gather*}
Note that the constant term $p_0$ is now excluded.
Consider the \emph{periodic Hilbert transform} $\mathcal H$,
\begin{gather*}
(\mathcal H\varphi)(z)=-\mathrm{i}\varphi_+(z)+\mathrm{i}\varphi_-(z)=\text{p.v.}\int_{S^1}
\frac{\mathrm{d}t}{2\pi}\cot\frac{s-t}{2}
\theta(t).
\end{gather*}
The periodic Benjamin--Ono equation is then characterized by the Hamiltonian
\begin{gather*}
\mathcal I=\int_{S^1}\frac{\mathrm{d}s}{2\pi}\left(\frac13\theta^{3}(s)+\frac{1}{2}
\theta^{\prime}(s)(\mathcal H\theta)(s)\right)
=\sum_{m,n\ge1}
(p_mp_n\bar{p}_{m+n}+p_{m+n}\bar{p}_m\bar{p}_n)-\sum_{n\ge1}np_n\bar{p}_n
\end{gather*}
which is simply the classical limit (as $\alpha\rightarrow0$) of the operator~\eqref{def-I}.
The correspondence rule between the quantum commutator and classical Poisson bracket we use here is
$\alpha^{-1}[\;,\; ]\to\mathrm{i}\{\;,\;\}$.

In the classical case, the equations of motion determined by the Hamiltonian $\mathcal I$ read
\begin{gather}
\label{BOeq}
\partial\varphi(z)/\partial t=\{\varphi(z)
,\mathcal I\}=\mathrm{i}(z\partial /\partial
z)^2(\varphi_+-\varphi_-)(z)-\mathrm{i}(z\partial /\partial z)\varphi^2(z),
\end{gather}
or equivalently,
\begin{gather}
\label{eqmo-p}
\partial p_n/\partial t=-\mathrm{i}n^2p_n+2\mathrm{i}n\sum_{m\ge1}p_{n+m}\bar{p}_m+\mathrm{i}n\sum_{1\le m<n}p_mp_{n-m},
\\
\label{eqmo-pp}
\partial \bar{p}_n/\partial t=
\mathrm{i}n^2\bar{p}_n-2\mathrm{i}n\sum_{m\ge1}\bar{p}_{n+m}{p_m}-\mathrm{i}n\sum_{1\le m<n}\bar{p}_m\bar{p}_{n-m}.
\end{gather}

The Benjamin--Ono equation is known to be a~completely integrable system having a~coun\-tab\-le set of higher
commuting Hamiltonians~\cite{AFSS,KLM,KM}.
The key to the integrability of the equation~\eqref{BOeq} is given by its {\it Lax representation}
\begin{gather}
\label{lax-pair-eq}
\partial \mathcal L/\partial t=[\mathcal M,\mathcal L],
\end{gather}
where the Lax operators $\mathcal L$ and $\mathcal M$ are the integro-dif\/ferential operators acting on
the functions
\begin{gather*}
f(z)=f_0+f_1z+f_2z^2+\cdots
\end{gather*}
and are def\/ined by
\begin{gather*}
\mathcal L:\ f\mapsto-z\partial f/\partial z+(\varphi f)_{+},
\\
\mathcal M:\ f\mapsto\mathrm{i}(z\partial /\partial z)^2f-2\mathrm{i}
\varphi_+z\partial f/\partial z-2\mathrm{i}
(z\partial (\varphi_-f)/\partial z)_{+}.
\end{gather*}

The functions $f(z)$ satisfying the condition $f(0)=0$ consitute an invariant subspace for the operators
$\mathcal L$ and $\mathcal M$ that is more convenient for us to use.
Identifying $f(z)$ with the column~vector
\begin{gather*}
\begin{bmatrix}
f_1
\\[2pt]
f_2
\\[6pt]
\raisebox{0pt}[0pt][2pt]{\vdots}
\end{bmatrix}
\end{gather*}
one can represent the operators $\mathcal L$ and $\mathcal M$ by the inf\/inite matrices as follows:
\begin{gather*}
\mathcal L=
\begin{bmatrix}
-1&p_1&p_2&p_3&\hspace{-1pt}\ldots
\\
\bar{p}_1&-2&p_1&p_2&\ddots
\\
\bar{p}_2&\bar{p}_1&-3&p_1&\ddots
\\[-2pt]
\raisebox{-1pt}{\vdots}\hspace{12pt}\ddots\hspace{-20pt}&\hspace{24pt}\ddots\hspace{-20pt}&\hspace{24pt}
\ddots\hspace{-20pt}&\hspace{24pt}\ddots\hspace{-20pt}
\end{bmatrix}
,
\qquad
\mathcal M=\mathrm{i}
\begin{bmatrix}
1&-2p_1&-2p_2&-2p_3&\hspace{-1pt}\ldots
\\
-2\bar{p}_1&4&-4p_1&-4p_2&\ddots
\\
-2\bar{p}_2&-4\bar{p}_1&9&-6p_1&\ddots
\\[-2pt]
\raisebox{-1pt}{\vdots}\hspace{12pt}\ddots\hspace{-20pt}&\hspace{24pt}\ddots\hspace{-20pt}&\hspace{24pt}
\ddots\hspace{-20pt}&\hspace{24pt}\ddots\hspace{-20pt}
\end{bmatrix}
.
\end{gather*}
It is easy to verify that the equation~\eqref{lax-pair-eq} for these matrices is equivalent
to~\eqref{eqmo-p} and~\eqref{eqmo-pp}.

The most common way to produce the commuting integrals of motion for an integrable system is to def\/ine
them as coef\/f\/icients of expansion in $u$ of the spectral determinant $\det(u-\mathcal
L)$, or equivalently, of its logarithmic derivative $\tr(u-\mathcal L)^{-1}$.
But our operator $\mathcal L$ is a~pertubation of the dif\/ferential operator
$-z\partial /\partial  z$ with a~Toeplitz matrix, and its spectrum is unbounded.
The above determinant and trace both diverge if understood literally.
Although it might be possible to regularize them, we shall take another approach by using a~peculiarity of
the Benjamin--Ono equation, namely the fact that it has a~rational spectral curve.

The integrable systems with a~rational spectral curve, for example the $N$-particle open Toda chain and the
Calogero--Moser model, possess peculiar properties that are missing for non-zero genus
spectral curves.
In particular, instead of $\tr(u-\mathcal
L)^{-1}$ one can use, as a~generating function of integrals of motion, the matrix element $r(u-\mathcal
L)^{-1} c$ for a~special pair $r$ and $c$ of row and column vectors.
See~\cite{KV,Skl,Ta} for the open Toda chain and~\cite{Ta,Va} for the Calogero--Moser~model.
In our case, the appropriate vectors are $r=p=[p_1\; p_2\; \ldots]$ and its conjugate
\begin{gather*}
c=\bar{p}=
\begin{bmatrix}
\bar{p}_1
\\[2pt]
\bar{p}_2
\\[6pt]
\raisebox{0pt}[0pt][2pt]{\vdots}
\end{bmatrix}
.
\end{gather*}

It is easy to show that the coef\/f\/icients of the generating function
\begin{gather}
\label{I(u)cl}
\mathcal I(u)=p(u-\mathcal L)^{-1}\bar{p}
\end{gather}
commute with the Hamiltonian $\mathcal I$ relative to the Poisson bracket.
Here the identity $\partial \mathcal I(u)/\partial  t=0$ follows immediately
from~\eqref{lax-pair-eq} and from the equalities
\begin{gather*}
\{\bar{p},\mathcal I\}=\mathcal M\bar{p}
\qquad
\text{and}
\qquad
\{p,\mathcal I\}=-p\mathcal M
\end{gather*}
which can be easily verif\/ied.
We do not discuss now the pairwise commutativity of all coef\/f\/icients of $\mathcal I(u)$ since we are
going to prove it in the quantum case, thus obtaining the commutativity in the classical case as
a~byproduct.

A natural question is whether there exists a~quantum analog of the formula~\eqref{I(u)cl}.
A strong indication that this might be the case is provided by the example of the $N$-particle quantum
Calogero--Moser system~\cite{UWH} where the quantum Lax matrix and the vectors $r$, $c$ coincide
with the classical ones.
We answer this question af\/f\/irmatively.
Namely, the quantum deformations of the Lax operator $\mathcal L$ and of the generating function $\mathcal
I(u)$ are given, respectively, by~\eqref{Lax_matrix}~and~\eqref{hua}.

\section{Monomial symmetric functions and powers sums}\label{Section3}

Fix any f\/ield $\mathbb F$.
Denote by $\Lambda$ the $\mathbb F$-algebra of symmetric functions in inf\/initely many variables
$x_1,x_2,\ldots$.
Following~\cite{M} we will introduce some standard bases of $\Lambda$.

Let $\lambda=(\lambda_1,\lambda_2,\ldots)$ be any partition of $0,1,2,\ldots$.
Here we assume that $\lambda_1\ge\lambda_2\ge\cdots$.
The number of non-zero parts is called the {\it length} of $\lambda$ and denoted by $\ell(\lambda)$.
We will also denote $|\lambda|=\lambda_1+\lambda_2+\cdots$ as usual.
Let $m_\lambda\in\Lambda$ be the \textit{monomial symmetric function} corresponding to the partition
$\lambda$.
By def\/inition,
\begin{gather}
\label{mon}
m_\lambda=\sum_{i_1<\dots<i_k}
\sum_{\sigma\in\mathfrak{S}_k}d_\lambda^{-1}x_{i_{\sigma(1)}}^{\lambda_1}\cdots x_{i_{\sigma(k)}}^{\lambda_k},
\end{gather}
where we write $k$ instead of $\ell(\lambda)$.
Here $\mathfrak{S}_k$ is the symmetric group permuting $1,\ldots, k$ while
\begin{gather}
\label{cela}
d_\lambda=k_1!k_2!\cdots
\end{gather}
if $k_1, k_2,\ldots$ are the respective multiplicites of the parts $1,2,\ldots$ of $\lambda$.
If $\ell(\lambda)=0$ then $m_\lambda=1$.
The monomial symmetric functions corresponding to partitions of $0,1,2,\ldots$ form a~basis of~$\Lambda$.

For $n=1,2,\ldots$ let $p_n\in\Lambda$ be the \textit{power sum symmetric function} of degree~$n$.
By def\/inition,
\begin{gather}
\label{pon}
p_n=x_1^n+x_2^n+\cdots.
\end{gather}
More generally, for any partition $\lambda$ put $ p_\lambda=p_{\lambda_1}\cdots p_{\lambda_k} $ where
$k=\ell(\lambda)$ as in~\eqref{mon}.
The ele\-ments~$p_\lambda$ form another basis of $\Lambda$.
In other words, the elements $p_1,p_2,\ldots$ are free generators of the commutative algebra~$\Lambda$ over
$\mathbb F$.
We will also use the formal power series in the variable $t$
\begin{gather*}
P(t)=p_1+p_2t+p_3t^2+\cdots.
\end{gather*}

Def\/ine a~bilinear form $\langle\;,\;\rangle$ on the vector space $\Lambda$ by setting for any two
partitions $\lambda$ and $\mu$
\begin{gather}
\label{schurprod}
\langle p_\lambda,p_\mu\rangle=z_\lambda\delta_{\lambda\mu},
\qquad
\text{where}
\qquad
z_\lambda=1^{k_1}k_1!2^{k_2}k_2!\cdots
\end{gather}
in the notation~\eqref{cela}.
This form is obviously symmetric and non-degenerate.
We will indicate by the superscript ${}^\perp$ the operator conjugation relative to this bilinear form.
In particular, by~\eqref{schurprod} for the operator conjugate to the multiplication in $\Lambda$ by $p_n$
with $n\ge1$ we have $ p_n^{\perp}=n\partial /\partial p_n.
{}$

The basis of $p_\lambda$ is related to the basis of monomial symmetric functions as follows.
For any two partitions $\lambda$ and $\mu$ let $r_{\lambda\mu}$ be the number of maps
$\theta:\{1,\ldots,\ell(\mu)\}\to\{1,2,\ldots\}$~such~that
\begin{gather}
\label{refine}
\sum_{\theta(j)=i}\mu_j=\lambda_i
\qquad
\text{for each}
\quad
i=1,2,\ldots.
\end{gather}
For any such $\theta$ the partition $\mu$ in~\eqref{refine} is called a~\textit{refinement} of $\lambda$.
Note that if $r_{\lambda\mu}\neq0$ then $|\lambda|=|\mu|$.
Moreover, then by [I.6.10] we have $\mu\le\lambda$ in the \textit{natural partial ordering} of partitions:
\begin{gather*}
\mu_1\le\lambda_1,\qquad \mu_1+\mu_2\le\lambda_1+\lambda_2,\qquad \ldots.
\end{gather*}
By [I.6.9] we have the relation
\begin{gather}\label{pm}
p_\mu=\sum_\mu r_{\lambda\mu}m_\lambda.
\end{gather}

\section{Elementary and complete symmetric functions}\label{Section4}

For $n=1,2,\ldots$ let $e_n\in\Lambda$ be the \textit{elementary symmetric function} of degree~$n$.
By def\/inition,
\begin{gather*}
e_n=\sum_{i_1<\dots<i_k}x_{i_1}\cdots x_{i_k}.
\end{gather*}
Note that by the def\/inition~\eqref{mon} we have $e_n=m_\lambda$ where $\lambda=(1^n)$.
Further, let $h_n\in\Lambda$ be the \textit{complete symmetric function} of degree $n$.
By def\/inition,
\begin{gather*}
h_n=\sum_{|\lambda|=n}m_\lambda.
\end{gather*}
We will also use the formal power series in the variable $t$
\begin{gather*}
E(t)=1+e_1t+e_2t^2+\cdots
\qquad
\text{and}
\qquad
H(t)=1+h_1t+h_2t^2+\cdots.
\end{gather*}
In particular, by [I.2.6] we have the relation
\begin{gather}
\label{eh}
E(-t)H(t)=1.
\end{gather}
Further, by [I.2.10] we have
\begin{gather}
\label{hep}
P(t)=E^{\prime}(-t)/E(-t).
\end{gather}

For any partition $\lambda$ denote $ e_\lambda=e_{\lambda_1}e_{\lambda_2}\cdots $ and $
h_\lambda=h_{\lambda_1}h_{\lambda_2}\cdots $ where $e_0=h_0=1$.
The elements $e_\lambda$ form a~basis of $\Lambda$, and so do the elements~$h_\lambda$.
It follows from~\eqref{eh} that for any~$n\ge1$
\begin{gather}
\label{ehnu}
(-1)^ne_n=\sum_{|\lambda|=n}(-1)^{\ell(\lambda)}h_\lambda\ell(\lambda)
!/d_\lambda.
\end{gather}
Furthermore, according to [I.4.5] the bases of $m_\lambda$ and $h_\lambda$ are dual to each other relative
to the bilinear form~\eqref{schurprod} on $\Lambda$:
\begin{gather}
\label{mh}
\langle m_\lambda,h_\mu\rangle=\delta_{\lambda\mu}.
\end{gather}
It follows from this duality that for any partitions $\lambda$ and $\nu$ we have
\begin{gather}
\label{hperp}
h_\nu^\perp(m_\lambda)=
\begin{cases}
m_\mu&\text{if}\quad\lambda=\mu\sqcup\nu,
\\
0&\text{otherwise,}
\end{cases}
\end{gather}
where $\mu\sqcup\nu$ is the partition obtained by collecting the parts of~$\mu$ and~$\nu$ together.
For~the~opera\-tor~$p_n^{\perp}$ with $n\ge1$ acting on the elements of~$\Lambda$ expressed as
polynomials in $h_1, h_2, \ldots$ by [Ex.~I.5.3]
\begin{gather}
\label{pperp}
p_n^{\perp}=\sum_{k\ge0}h_k\partial /\partial h_{n+k}.
\end{gather}

\section{Jack symmetric functions}\label{Section5}

Now let $\mathbb F$ be the f\/ield $\mathbb Q(\alpha)$ where $\alpha$ is another variable.
Generalizing~\eqref{schurprod} def\/ine a~bilinear form $\langle\;,\;\rangle_\alpha$ on $\Lambda$
by setting for any partitions $\lambda$ and $\mu$
\begin{gather}
\label{jackprod}
\langle p_\lambda,p_\mu\rangle_\alpha=\alpha^{\ell(\lambda)}
z_\lambda\delta_{\lambda\mu}.
\end{gather}
This form is symmetric and non-degenerate.
In particular, if $\alpha=1$ then this is the form $\langle\;,\; \rangle$.
We will indicate by the superscript ${}^\ast$ the operator conjugation relative to the form
$\langle\;,\;\rangle_\alpha$.
Using the def\/inition~\eqref{jackprod} for the operator conjugate to the multiplication in $\Lambda$ by
$p_n$ with $n\ge1$ we~get
\begin{gather}
\label{pnast}
p_n^{\ast}=\alpha n\partial /\partial p_n.
\end{gather}

By [Ex.~VI.4.2] there exists a~unique family of elements $J_\lambda\in\Lambda$ such that
\begin{gather}
\label{plamu}
\langle J_\lambda,J_\mu\rangle_\alpha=0
\qquad
\text{for}
\quad
\lambda\neq\mu
\end{gather}
and such that the $J_\lambda$ equals $m_\lambda$ plus a~linear combination of the elements $m_\mu$ with
$\mu<\lambda$ in the natural partial ordering.
The elements $J_\lambda\in\Lambda$ are called the \textit{Jack symmetric functions}.
In the case $\alpha=1$ the $J_\lambda$ coincides with the \emph{Schur symmetric function}~$s_\lambda$.

The Jack symmetric functions $J_\lambda$ form a~basis of the vector space $\Lambda$ over the f\/ield
$\mathbb Q(\alpha)$.
Let us now def\/ine the operators $A^{(1)},A^{(2)},\ldots$ acting on
$\Lambda$ as follows.
Put
\begin{gather}
\label{au}
A(u)=1+A^{(1)}/(u)_1+A^{(2)}/(u)_2+\cdots,
\end{gather}
where $u$ is another variable and $(u)_k=u(u+1)\cdots(u+k-1)$ is the \emph{Pochhammer symbol}.
Then
\begin{gather}
\label{ajack}
A(u)J_\lambda=\prod_{i\ge1}\frac{u+i-1-\alpha\lambda_i}{u+i-1}\cdot J_\lambda
\end{gather}
by def\/inition.
Note that in the inf\/inite product displayed above the only factors dif\/ferent from~$1$ are those
corresponding to $i=1,\ldots,\ell(\lambda)$.
In our recent publication~\cite{NS1} we proved the following theorem.
Another proof can be obtained by using the results of~\cite{Shi}, see also~\cite{NS2}.
\begin{theorem}\label{Theorem1}
In the notation~\eqref{cela} for each $k=1,2,\ldots$ we have
\begin{gather}
\label{basic}
A^{(k)}=(-1)^k\sum_{\ell(\lambda)=k}d_\lambda m_\lambda m_\lambda^{\ast},
\end{gather}
where the sum is taken over all partitions $\lambda$ of the fixed length $k$.
\end{theorem}

The Jack symmetric functions are eigenvectors of the operators
$A^{(1)},A^{(2)},\ldots$ by def\/inition.
In particular, these operators are self-conjugate relative to the form $\langle\;,\;\rangle_\alpha$ by the
orthogonality condition~\eqref{plamu}.
This self-conjugacy is also transparent from the explicit formula~\eqref{basic}.

By inverting the relation~\eqref{pm} any monomial symmetric function $m_\lambda$ can be expressed as
a~linear combination of the functions $p_\mu$ where $\lambda$, $\mu$ are partitions of the same number and~$\lambda\le\mu$.
By substituting into~\eqref{basic} and using~\eqref{pnast}, one can write each operator
$A^{(k)}$ in terms of~$p_n$ and $\partial /\partial  p_{n}$ where $n=1,2,\ldots$.
However, the result of this procedure for arbitrary $k$ does not seem to be suf\/f\/iciently explicit.
In the next section we will overcome this dif\/f\/iculty by considering another collection of operators on
$\Lambda$ such that the Jack symmetric functions are their eigenvectors.

\section{Lax matrix}\label{Section6}

Let us def\/ine the operators $I^{(1)},I^{(2)},\ldots$ acting on the vector
space $\Lambda$ over $\mathbb Q(\alpha)$ by the equations~\eqref{iu} and~\eqref{hdef}.
In particular, then we have the relations
\begin{gather*}
I^{(1)}=-A^{(1)}
\qquad
\text{and}
\qquad
I^{(2)}=A^{(1)}\big(A^{(1)}
+1\big)-2A^{(2)}.
\end{gather*}
Note that then due to the def\/inition~\eqref{ajack} of $A(u)$ we also have the equality
\begin{gather*}
I(u)J_\lambda=uJ_\lambda-(u+\ell(\lambda))\prod_{1\le i
\le\ell(\lambda)}\frac{u+i-1-\alpha\lambda_i}{u+i-\alpha\lambda_i}\cdot J_\lambda.
\end{gather*}

Let us now introduce the inf\/inite matrix
\begin{gather}
\label{Lax_matrix}
L=
\begin{bmatrix}
(\alpha-1)&p_1&p_2&p_3&\hspace{-1pt}\ldots
\\
p_1^*&2(\alpha-1)&p_1&p_2&\ddots
\\
p_2^*&p_1^*&3(\alpha-1)&p_1&\ddots
\\[-2pt]
\raisebox{-1pt}{\vdots}\hspace{12pt}\ddots\hspace{-20pt}&\hspace{24pt}\ddots\hspace{-20pt}
&\hspace{24pt}\ddots\hspace{-20pt}&\hspace{24pt}\ddots\hspace{-20pt}
\end{bmatrix},
\end{gather}
where the rows and columns can be labelled by the indices $i,j=1,2,\ldots$.
We shall call it the \emph{Lax matrix}.
It is convenient to set $p_{-n}=p^{\ast}_n$ for $n\ge1$.
We also keep assuming that $p_{0}=0$.
Then $L=[ L_{ij}]_{i,j=1}^{\infty}$ where
\begin{gather}
\label{Ljk}
L_{ij}=j(\alpha-1)\delta_{ij}+p_{j-i}.
\end{gather}
Note that the matrix $L$ is self-conjugate relative to the form $\langle\; ,\; \rangle_\alpha$.
Also introduce the inf\/inite row vector $p=[p_1\; p_2\; \ldots]$ and its conjugate column vector
\begin{gather*}
p^{\ast}=
\begin{bmatrix}
p^{\ast}_1
\\[2pt]
p^{\ast}_2
\\[6pt]
\raisebox{0pt}[0pt][2pt]{\vdots}
\end{bmatrix}
.
\end{gather*}
The main result of the present article is the next theorem which provides an explicit expression for every
operator $I^{(k)}$ in terms of $p_n$ and $\partial /\partial  p_{n}$ where
$n=1,2,\ldots$.
\begin{theorem}\label{Theorem2}
We have the equality
\begin{gather}
\label{hua}
I(u)=p(u-L)^{-1}p^{\ast},
\end{gather}
where the inverse to the infinite matrix $u-L$ is regarded as formal power series in $u^{-1}$:
\begin{gather*}
(u-L)^{-1}=u^{-1}+Lu^{-2}+L^2u^{-3}+\cdots.
\end{gather*}
\end{theorem}

Note that each coef\/f\/icient of the series in $u^{-1}$ at the right hand side of the
equality~\eqref{hua} is an inf\/inite sum of operators acting on $\Lambda$.
However, only f\/inite number of the summands do not vanish on any subspace in $\Lambda$ of a~f\/ixed
degree in $x_1,x_2,\ldots$.
This can be easily seen by restating Theorem~\ref{Theorem2} as the equality for every index $k\ge1$
\begin{gather}
\label{hk}
I^{(k)}=\sum_{i_1,\ldots,i_k=1}^\infty p_{i_1}
L_{i_1i_2}\cdots L_{i_{k-1}i_k}p^{\ast}_{i_k}
\end{gather}
and by performing summation in~\eqref{hk} consecutively over the indices $i_k,i_{k-1},\ldots,
i_1$.
In the next section we explain our method of proving Theorem~\ref{Theorem2}.
The proof will be completed after that.

\section[Baker-Akhiezer vector]{Baker--Akhiezer vector}\label{Section7}

Consider the inf\/inite column vector
\begin{gather}
\label{psiu}
\Psi(u)=
\begin{bmatrix}
\raisebox{0pt}[12pt][0pt]{$\Psi_1(u)$}
\\
\Psi_2(u)
\\
\vdots
\end{bmatrix}
=(u-L)^{-1}p^{\ast}A(u+1).
\end{gather}
Every entry $\Psi_n(u)$ of this vector is a~formal power series in $u^{-1}$ with coef\/f\/icients acting on
$\Lambda$; see the remark in the end of previous section.
We shall call it the \emph{Baker--Akhiezer vector} for the Lax matrix~\eqref{Lax_matrix}, since in the
classical limit this is an analog of the Baker--Akhiezer function for a~rational spectral curve~\cite{Skl}.
Our proof of Theorem~\ref{Theorem2} is based on the explicit formula for $\Psi_n(u)$ given by the next theorem.
This theorem is another principal result of the present article.
Let
\begin{gather*}
\Psi_n(u)=\Psi^{(1)}_n/(u+1)_1+\Psi^{(2)}_n/(u+1)_2+\cdots,
\end{gather*}
where the coef\/f\/icients $\Psi^{(1)}_n,\Psi^{(2)}_n,\ldots$ are operators
acting on the vector space $\Lambda$ over $\mathbb Q(\alpha)$.
\begin{theorem}\label{Theorem3}
For any indices $k,n\ge1$ we have the equality
\begin{gather}
\label{psink}
\Psi^{(k)}_n=(-1)^{n+k}\sum_{\ell(\lambda)=k}d_\lambda e_n^{
\perp}(m_\lambda)m_\lambda^{\ast}.
\end{gather}
\end{theorem}

Note that by using the relation~\eqref{ehnu} and then~\eqref{hperp} here we can write
\begin{gather}\label{firstfactor}
(-1)^ne_n^{\perp}(m_\lambda)
=\sum\limits_{\substack{\lambda=\mu\sqcup\nu\\|\nu|=n}}(-1)^{\ell(\nu)}m_\mu\ell(\nu)!/d_\nu,
\end{gather}
where $\nu$ ranges over all partions of $n$ such that by taking the parts of $\nu$ together with those of
another partition $\mu$ we get $\lambda$.
The partition $\mu$ is then determined by $\lambda$ and $\nu$ uniquely.

Theorem~\ref{Theorem3} shows that the series $\Psi_n(u)$ can be obtained from the series $A(u+1)$ by applying the
operator $(-1)^ne_n^{\perp}$ to the factor $m_\lambda$ in the
coef\/f\/icient~\eqref{basic}.
Thus $\Psi_n(u)$ can be expressed in terms of $A(u+1)$ without using Theorem~\ref{Theorem1}.
However we will prove Theorem~\ref{Theorem3} as stated here.

Theorem~\ref{Theorem2} will then follow rather easily.
Indeed, due to the def\/inition~\eqref{psiu} of $\Psi(u)$ to obtain the stated equality~\eqref{hua} it
suf\/f\/ices to prove that
\begin{gather*}%\label{pha}
p\Psi(u)=I(u)A(u+1).
\end{gather*}
By using the def\/inition~\eqref{hdef} of $I(u)$ the last displayed equality can be rewritten as
\begin{gather}
\label{pepsi}
p\Psi(u)=uA(u+1)-uA(u).
\end{gather}
Here on the left hand side we have a~product of a~row vector by a~column vector:
\begin{gather*}
p\Psi(u)=\sum_{n\ge1}p_n\Psi_n(u).
\end{gather*}
By using Theorems~\ref{Theorem1} and~\ref{Theorem3} along with the obvious relation for any $k\ge1$
\begin{gather*}
u/(u+1)_k-u/(u)_k=-k/(u+1)_k,
\end{gather*}
the equality~\eqref{pepsi} follows from the proposition which is
stated and proved below.
\begin{proposition}\label{Prop1}
For any partition $\lambda$ of the length $k\ge1$ we have
\begin{gather}\label{pro1}
\sum_{n\ge1}(-1)^np_ne_n^{\perp}(m_\lambda)=-km_\lambda.
\end{gather}
\end{proposition}
\begin{proof}
Using the bilinear form $\langle\;,\;\rangle$ take the operator on $\Lambda$ conjugate to the operator which
is applied to $m_\lambda$ at the left hand side of the equality~\eqref{pro1}.
The conjugate operator equals
\begin{gather*}
\sum_{n\ge1}(-1)^ne_np_n^{\perp}
=\sum_{n\ge1}\sum_{k\ge0}(-1)^ne_nh_k\partial /\partial h_{n+k}
\\
\phantom{\sum_{n\ge1}(-1)^ne_np_n^{\perp}}
=\sum_{n\ge1}\sum_{0\le k<n}(-1)^{n-k}e_{n-k}
h_k\partial /\partial h_n=-\sum_{n\ge1}h_n\partial /\partial h_n,
\end{gather*}
where we used the relation~\eqref{pperp} and then~\eqref{eh}.
Therefore by applying the conjugate operator to the vector $h_\lambda$ with $\ell(\lambda)=k$ we get
$- kh_\lambda$.
Now~\eqref{pro1} follows from the duality relation~\eqref{mh}.
\end{proof}

\section{Proof of Theorem~\ref{Theorem3}}\label{Section8}

Denote by $\Phi^{(k)}_n$ the right hand side of the relation~\eqref{psink}.
Then introduce the~series
\begin{gather}
\label{phin}
\Phi_n(u)=\Phi^{(1)}_n/(u+1)_1+\Phi^{(2)}_n/
(u+1)_2+\cdots
\end{gather}
and the column vector
\begin{gather*}
\Phi(u)=
\begin{bmatrix}
\raisebox{0pt}[12pt][0pt]{$\Phi_1(u)$}
\\
\Phi_2(u)
\\
\vdots
\end{bmatrix}
.
\end{gather*}
We have to prove the relation $\Psi(u)=\Phi(u)$.
Multiplying by $u-L$ in the def\/inition~\eqref{psiu} of $\Psi(u)$ the latter relation becomes
\begin{gather*}
p^{\ast}A(u+1)=(u-L)\Phi(u).
\end{gather*}
By taking the entries of the column vectors here and using the def\/inition~\eqref{Ljk} we have to prove
that for every index $n\ge1$
\begin{gather}
\label{pjA}
p^{\ast}_nA(u+1)=(u+n-\alpha n)\Phi_n(u)-\sum_{i\ge1}
p_{i-n}\Phi_i(u).
\end{gather}

By using for $n\ge1$ the expansions~\eqref{au} and~\eqref{phin} along with the obvious relation for $k\ge1$
\begin{gather*}
u/(u+1)_k=1/(u+1)_{k-1}-k/(u+1)_k,
\end{gather*}
the relation~\eqref{pjA} is equivalent to the collection of relations for all $k\ge1$
\begin{gather}
\label{phink}
p^{\ast}_nA^{(k)}=\Phi^{(k+1)}_n+(
n-\alpha n-k)\Phi^{(k)}_n-\sum_{i\ge1}
p_{i-n}\Phi^{(k)}_n
\end{gather}
together with the relation $p^{\ast}_n=\Phi^{(1)}_n$.
But the latter relation follows from the def\/inition of $\Phi^{(1)}_n$ by
using~\eqref{firstfactor}.
Indeed, if $\lambda=\mu\sqcup\nu$ where $\ell(\lambda)=1$ and $|\nu|=n\ge1$, then $\lambda=\nu=(n)$ while
$\mu$ has no parts dif\/ferent from zero.
We then observe that $p_n^{\ast}=m_{(n)}^{\ast}$.
So we have to prove~\eqref{phink}.

For any two symmetric functions $f,g\in\Lambda$ the operator $f g^{\ast}$
acting on the vector space $\Lambda$ can be represented by its \emph{symbol} $f\otimes
g\in\Lambda\otimes\Lambda$.
We will extend this representation to inf\/inite sums of operators of the form $f
g^{\ast}$ by linearity.
In particular, the symbol of the operator at the right hand side of~\eqref{basic} is the inf\/inite linear
combination of elements of the vector space $\Lambda\otimes\Lambda$
\begin{gather}
\label{dek}
\Delta^{(k)}=(-1)^k\sum_{\ell(\lambda)=k}d_\lambda m_\lambda\otimes m_\lambda.
\end{gather}

Extending our notation, set $\Phi^{(k)}_0=A^{(k)}$.
By Theorem~\ref{Theorem1} this is indeed the right hand side of~\eqref{psink} at $n=0$.
Thus for any $n\ge0$ the symbol of the operator $\Phi^{(k)}_n$ equals $(-1)^{
n}(e_n^{\perp}\otimes1)\Delta^{(k)}$.

Let us now prove~\eqref{phink} for $n,k\ge1$.
Using our extended notation,~\eqref{phink} can be rewritten~as
\begin{gather*}
\sum_{1\le i\le n}p^{\ast}_i\Phi_{n-i}^{(k)}+\sum_{i\ge1}
p_i\Phi_{n+i}^{(k)}=\Phi^{(k+1)}_n+(
n-\alpha n-k)\Phi_n^{(k)}.
\end{gather*}
In the representation by symbols the latter operator relation becomes
\begin{gather}
\sum_{1\le i\le n}(-1)^{n-i}\big(\alpha p^{\perp}_i e_{n-i}^{\perp}\otimes1+e_{n-i}^{\perp}\otimes p_i\big)\Delta^{(k)}
+\sum_{i\ge1}(-1)^{n+i}\big(p_ie_{n+i}^{\perp}\otimes1\big)\Delta^{(k)}
\nonumber
\\
\qquad {}
=(-1)^{n}\big(e_n^{\perp}\otimes1\big)\Delta^{(k+1)}+(n-\alpha n-k)(-1)^{n}\big(e_n^{\perp}\otimes1\big)\Delta^{(k)}.
\label{symb}
\end{gather}
Here we take commutators with $p^{\ast}_i$ and then use the equality
$p^{\ast}_i=\alpha p^{\perp}_i$ given by~\eqref{pnast}.

Observe that both sides of the relation~\eqref{symb} are polynomials in $\alpha$ of degree
$1$.
By equating their coef\/f\/icients at~$\alpha$ we get the relation
\begin{gather*}
\sum_{1\le i\le n}(-1)^{n-i}\big(p^{\perp}_i e_{n-i}
^{\perp}\otimes1\big)\Delta^{(k)}=n(-1)^{n+1}
\big(e_n^{\perp}\otimes1\big)\Delta^{(k)}.
\end{gather*}
But this relation follows from the identity
\begin{gather*}
\sum_{1\le i\le n}(-1)^{n-i}e_{n-i}p_i=n(-1)^{n+1}e_n,
\end{gather*}
which holds by~\eqref{hep}.
It now remains to verify the relation~\eqref{symb} only for $\alpha=0$, that is
\begin{gather}
\sum_{1\le i\le n}(-1)^{n-i}\big(e_{n-i}^{\perp}
\otimes p_i\big)\Delta^{(k)}+\sum_{i\ge1}(-1)^{n+i}
\big(p_ie_{n+i}^{\perp}\otimes1\big)\Delta^{(k)}
\nonumber
\\
\qquad{}
=(-1)^{n}\big(e_n^{\perp}\otimes1\big)\Delta^{(k+1)}+(
n-k)(-1)^{n}\big(e_n^{\perp}\otimes1\big)\Delta^{(k)}.
\label{symbzero}
\end{gather}

It follows from the def\/initions~\eqref{mon} and~\eqref{pon} that for any integer $i\ge1$ and any
partition $\mu$
\begin{gather}\label{pim}
p_im_\mu=\sum_\lambda c_{\lambda\mu}m_\lambda,
\end{gather}
where $\lambda$ ranges over all those partitions which can be obtained by increasing by $i$ anyone of the
\emph{distinct} parts of $\mu$.
That includes increasing a~zero part of $\mu$, in which case $\ell(\lambda)=\ell(\mu)+1$.
Otherwise we have $\ell(\lambda)=\ell(\mu)$ of course.
In any case, the coef\/f\/icient $c_{\lambda\mu}$ in~\eqref{pim} is equal to the
multiplicity in $\lambda$ of that part which has been obtained by increasing.
Note that if the increased part of $\mu$ is zero, that is if $\lambda=\mu\sqcup i$, then
the product $c_{\lambda\mu}d_\mu$ equals $d_\lambda$.
But if the increased part of $\mu$ is not zero, then the ratio $
c_{\lambda\mu}d_\mu/ d_\lambda $ equals the multiplicity of
the increased part of $\mu$.
Let us denote the latter multiplicity by $c_{\mu\lambda}$, so that we have
$c_{\lambda\mu}d_\mu=c_{\mu\lambda}d_\lambda$ when
$\ell(\lambda)=\ell(\mu)$.

The expansion~\eqref{pim} also shows that either side of the relation~\eqref{symbzero} can be written as
a~sum of tensor products of the form $f\otimes m_\lambda$ where $f\in\Lambda$ and $\ell(\lambda)=k,k+1$;
see the def\/inition~\eqref{dek}.
By taking only the summands $f\otimes m_\lambda$ in~\eqref{symbzero} where
$\ell(\lambda)=k+1$, we get the relation
\begin{gather*}
\sum_{1\le i\le n}(-1)^{n-i+k}\sum_{\ell(\mu)=k}c_{\mu
\sqcup i,\mu}d_\mu e_{n-i}^{\perp}(m_\mu)\otimes m_{\mu\sqcup i}
=(-1)^{n+k+1}\sum_{\ell(\lambda)=k+1}d_\lambda e_n^{\perp}(m_\lambda)\otimes m_\lambda.
\end{gather*}
By using~\eqref{hperp} and the equality $ c_{\mu\sqcup
i,\mu}d_\mu=d_{\mu\sqcup i}
$ the last displayed relation be rewritten as
\begin{gather*}
\sum_{1\le i\le n}(-1)^{n-i+k}\sum_{\ell(\lambda)=k+1}d_\lambda e_{n-i}
^{\perp}h_i^{\perp}(
m_\lambda)\otimes m_\lambda=(-1)^{n+k+1}\sum_{\ell(\lambda)=k+1}
d_\lambda e_n^{\perp}(m_\lambda)\otimes m_\lambda.
\end{gather*}
But this relation immediately follows from an identity which holds by~\eqref{eh},
\begin{gather*}
\sum_{1\le i\le n}(-1)^{n-i}h_ie_{n-i}=(-1)^{n+1}e_n.
\end{gather*}

Now take the summands $f\otimes m_\lambda$ in~\eqref{symbzero} where $\ell(\lambda)=k$.
In this way we get the relation
\begin{gather}
\sum_{1\le i\le n}(-1)^{n-i+k}\sum_{\ell(\mu)=k}\sum_\lambda
\qquad
c_{\lambda\mu}d_\mu e_{n-i}^{\perp}(m_\mu)\otimes m_\lambda
\nonumber
\\
\qquad\phantom{=}
{}+\sum_{i\ge1}\sum_{\ell(\lambda)=k}(-1)^{n+i+k}d_\lambda
p_ie_{n+i}^{\perp}(m_\lambda)\otimes m_\lambda
\nonumber
\\
\qquad{}
=(n-k)(-1)^{n+k}\sum_{\ell(\lambda)=k}d_\lambda e_n^{\perp}(m_\lambda)\otimes m_\lambda,
\label{threeline}
\end{gather}
where in the f\/irst of the three displayed lines the partition $\lambda$ is obtained by increasing by $i$
any of the distinct non-zero parts of $\mu$.
Let us consider the sum in the second line of the display~\eqref{threeline}.
\begin{proposition}\label{Prop2}
For any $n\ge1$ we have the operator identity
\begin{gather}
\label{prop2}
\sum_{i\ge1}(-1)^{n+i}p_ie_{n+i}^{\perp}
=\sum_{0\le i\le n}\sum_{j\ge1}(-1)^{n-i+1}(
\partial /\partial h_j)^{\perp}e_{n-i}
^{\perp}h_{i+j}^{\perp}.
\end{gather}
\end{proposition}
\begin{proof}
Using the form $\langle\;,\;\rangle$ let us conjugate the operator at the left hand side of~\eqref{prop2}.
We get
\begin{gather*}
\sum_{i\ge1}(-1)^{n+i}e_{n+i}p_i^{\perp}
=\sum_{i\ge1}\sum_{l\ge0}(-1)^{n+i}e_{n+i}h_{l}\partial /\partial h_{i+l}
=\sum_{j\ge1}\sum_{0\le l<j}(-1)^{n+j-l}e_{n+j-l}h_{l}\partial /\partial h_{j}
\\
\phantom{\sum_{i\ge1}(-1)^{n+i}e_{n+i}p_i^{\perp}}
=\sum_{j\ge1}\sum_{j\le l\le n+j}(-1)^{n+j-l+1}e_{n+j-l}h_{l}\partial /\partial h_{j}
\\
\phantom{\sum_{i\ge1}(-1)^{n+i}e_{n+i}p_i^{\perp}}
=\sum_{j\ge1}\sum_{0\le i\le n}(-1)^{n-i+1}e_{n-i}h_{i+j}\partial /\partial h_{j}
\end{gather*}
as required.
Here we use~\eqref{pperp}, then denote $j=i+l$, then use~\eqref{eh}, and f\/inally set
$i=l-j$.
\end{proof}

Further, the right hand side of the identity~\eqref{prop2} can be rewritten as
\begin{gather*}
\sum_{0\le i\le n}\sum_{j\ge1}(-1)^{n-i+1}e_{n-i}^{\perp}(\partial /\partial h_j)^{\perp}
h_{i+j}^{\perp}+\sum_{0\le i\le n}\sum_{j\ge1}(-1)^{n-i}(\partial e_{n-i}/\partial h_j)^{\perp}h_{i+j}^{\perp},
\end{gather*}
where we use the commutator of the operator $(\partial /\partial
h_j)^{\perp}$ with $e_{ n-i}^{\perp}$.
Consider the conjugate of the second group of summands in the above display.
The conjugate is an operator of multiplication by the element of $\Lambda$ which is computed in the next
proposition.
\begin{proposition}\label{Prop3}
For any $n\ge1$ we have the identity in $\Lambda$
\begin{gather}
\label{prop3}
\sum_{0\le i\le n}\sum_{j\ge1}(-1)^{n-i}h_{i+j}
\partial e_{n-i}/\partial
h_j=(-1)^nne_n.
\end{gather}
\end{proposition}
\begin{proof}
The summation at the left hand side of~\eqref{prop3} can be restricted to $0\le i<n$~without af\/fecting
the value of the sum, because $e_0=1$.
Then by using~\eqref{ehnu} the sum can be rewritten as
\begin{gather}
\sum_{0\le i<n}\sum_{j\ge1}h_{i+j}\sum_{|\mu|=n-i}(-1)^{\ell(\mu)}
\partial h_\mu/\partial h_j\cdot\ell(\mu)
!/d_\mu
\nonumber
\\
\qquad\qquad
=\sum_{0\le i<n}\sum_{j\ge1}\sum_{|\mu|=n-i}(-1)^{\ell(\mu)}h_\nu\cdot c_{\mu\nu}\ell(\mu)!/d_\mu,
\label{prop4}
\end{gather}
where $\mu$ ranges over all partitions of $n-i$ containing $j$ as a~part.
Further, here $\nu$ denotes the partition obtained from $\mu$ by replacing a~part $j$ by $i+j$.
According to the notation introduced after stating~\eqref{pim}, here $c_{\mu\nu}$ is the
multiplicity of the part $j$ in $\mu$.
In particular, $\ell(\mu)=\ell(\nu)$ and
\begin{gather}
\label{cd}
c_{\mu\nu}/d_\mu=c_{\nu\mu}
/d_\nu.
\end{gather}
Here \looseness=1 we extend the notation $c_{\mu\nu}$ to the case $i=0$.
Then $\mu=\nu$ and the part $j$ is not determined by the notation.
This should not cause us any confusion however, as it will always remain clear which multiplicity is taken.
In particular, the equality~\eqref{cd} will remain valid in the case $i=0$.

By again using~\eqref{ehnu} the sum at the right hand side of~\eqref{prop3} can be rewritten as
\begin{gather}
\label{prop5}
\sum_{|\nu|=n}(-1)^{\ell(\nu)}h_\nu\cdot n\ell(\nu)!/d_\nu
\end{gather}
where \looseness=1 $\nu$ ranges over all partitions of $n$.
To prove the identity~\eqref{prop3} it remains to observe that for any f\/ixed $\nu$ the terms at the right
hand side of~\eqref{prop4} containing the factor $h_\nu$ can be collected as follows.
Choose any of the distinct non-zero parts of $\nu$ and replace it by a~positive integer not exceeding the
original part, hence getting some $\mu$.
For any choice of the original part there are as many possible replacements as the size of the part.
By multiplying the size by the ratio~\eqref{cd} and then summing over all distinct non-zero parts of $\nu$
we get $n/d_\nu$, exactly as in~\eqref{prop5}.
\end{proof}

We can now complete our proof of Theorem~\ref{Theorem3}.
By using Proposition~\ref{Prop2} and then Proposition~\ref{Prop3}
as explained just before stating the latter, the sum in the
second line of the display~\eqref{threeline} equals
\begin{gather}
\sum_{0\le i\le n}\sum_{j\ge1}\sum_{\ell(\lambda)=k}(-1)^{n-i+k+1}
d_\lambda e_{n-i}^{\perp}(\partial /\partial h_j)^{\perp}h_{i+j}^{\perp}(m_\lambda)\otimes m_\lambda
\nonumber
\\
\qquad\qquad
{}+\sum_{\ell(\lambda)=k}(-1)^{n+k}ne_n^{\perp}(m_\lambda)\otimes m_\lambda.
\label{last}
\end{gather}
In \looseness=1 the f\/irst line of the above display the summation can be restricted to partitions $\lambda$ containing
$i+j$ as a~part, without af\/fecting the value of the sum; see~\eqref{hperp}.
For these partitions $\lambda$ we have
\begin{gather*}
d_\lambda(\partial /\partial h_j)^{\perp}
h_{i+j}^{\perp}(m_\lambda)=d_\lambda c_{\mu\lambda}m_\mu=d_\mu c_{\lambda\mu}m_\mu,
\end{gather*}
where $\mu$ denotes the partition obtained from $\lambda$ by replacing a~part $i+j$ with $j$; see the
remark we made immediately after displaying~\eqref{cd}.
Therefore the summands in~\eqref{last} with $1\le i\le n$ cancel with the sum in the f\/irst line
of~\eqref{threeline}.
Since the sum of the multiplicities of non-zero parts of $\lambda$ is $\ell(\lambda)$, the summands
in~\eqref{last} with $i=0$ add up to
\begin{gather*}
\sum_{\ell(\lambda)=k}(-1)^{n+k+1}d_\lambda ke_n^{\perp}(m_\lambda)\otimes m_\lambda.
\end{gather*}
The \looseness=1 latter sum and the sum in the second line of~\eqref{last} cancel with the right hand side
of~\eqref{threeline}.
All these cancellations establish the relation~\eqref{threeline}.
Our proof of Theorem~\ref{Theorem3} is now completed.

\subsection*{Acknowledgements}

We are very grateful to S.M.~Khoroshkin and J.~Shiraishi for friendly discussions.
We cannot praise enough the wonderful Maple packages by J.~Stembridge~\cite{St} that facilitated our work
with symmetric functions.
The f\/irst and the second named of us have been supported by the EPSRC grants
EP/I014071 and EP/H000054 respectively.

\pdfbookmark[1]{References}{ref}
\LastPageEnding

\end{document}